\documentclass[ams,pra,twocolumn,amssymb,superscriptaddress,longbibliography,floatfix]{revtex4-2}
\usepackage{amsmath}
\usepackage{tabularx}
\usepackage{bm}
\usepackage{slashed}
\usepackage{euscript}
\usepackage{graphicx}
\usepackage{xcolor}
\usepackage{amsfonts}
\usepackage{exscale}
\usepackage{amsbsy}
\usepackage{subfigure}
\usepackage{textcomp}
\usepackage{comment}
\usepackage{mathrsfs}
\usepackage{afterpage}

\usepackage[normalem]{ulem} 

\usepackage{bbold}
\usepackage{blkarray}

\usepackage{ulem}

 \usepackage{hyperref}

\newcommand{\be}{\begin{equation}}
\newcommand{\ee}{\end{equation}}

\newcommand{\ep}{\varepsilon}

\begin{document}

\title{Interplay of superconductivity and localization near a 2D ferromagnetic quantum critical point} 

	\author{P. A. Nosov}
	
	\affiliation{Stanford Institute for Theoretical Physics, Stanford University, Stanford, California 94305, USA}
	
	\author{I. S. Burmistrov}
	\affiliation{L.D. Landau Institute for Theoretical Physics, acad. Semenova av.1-a, 142432, Chernogolovka, Russia}
	\affiliation{\hbox{Laboratory for Condensed Matter Physics, HSE University, 101000 Moscow, Russia}}
	
	\author{S. Raghu}
	
	\affiliation{Stanford Institute for Theoretical Physics, Stanford University, Stanford, California 94305, USA}
	
	\affiliation{\hbox{Stanford Institute for Materials and Energy Sciences, SLAC National Accelerator Laboratory, Menlo Park, California 94025, USA}}
	\date{\today}
	
	\begin{abstract}
We study the superconducting instability of a two-dimensional 
disordered Fermi liquid weakly coupled to the soft fluctuations associated with proximity to an Ising-ferromagnetic quantum critical point. We derive interaction-induced corrections to the Usadel equation governing the superconducting gap function, and show that diffusion and localization effects drastically modify the interplay between fermionic incoherence and strong pairing interactions. In particular, we obtain the phase diagram, and demonstrate that: (i) there is an intermediate range of disorder strength where superconductivity is enhanced, eventually followed by a tendency towards the superconductor-insulator transition at stronger disorder; and (ii) diffusive particle-particle modes (so-called ``Cooperons") acquire anomalous dynamical scaling $z=4$, indicating strong non-Fermi liquid behavior.
	\end{abstract}

	\maketitle
\section{Introduction}
 In a variety of strongly correlated materials, superconductivity develops out of a normal state without well-defined quasiparticles.  Perhaps the best example of such phenomena is the cuprate family, which exhibit high temperature superconductivity optimally enhanced in the vicinity of an apparent quantum critical point accessed by doping\cite{Broun2008,Proust2019}.  Other examples include the iron-based superconductors \cite{Stewart2011,PhysRevLett.104.057006,Nematic2017}, and a host of heavy fermion materials where magnetic quantum critical fluctuations appear to enhance superconducting tendencies \cite{mathur1998magnetically,saxena2000superconductivity}.  
The interplay between pairing and non-Fermi liquid (NFL) behavior at two-dimensional (2D) metallic quantum critical points (QCPs) is often invoked to explain superconductivity born out of incoherent quasiparticles \cite{combescot1995strong,bonesteel1996gauge,son1999superconductivity,abanov2001coherent,abanov2003quantum,lee2009low,moon2010quantum,mahajan2013quantum,metlitski2015cooper,lederer2015enhancement,Ipsita,wang2017non,wang2018superconductivity,abanov2020interplay,wu2020interplay}.  The same soft order parameter fluctuations that enhance superconductivity act  also to destroy Landau quasiparticles, rendering them incoherent and opposing the trend towards superconductivity.    Thus in principle, there are several logically distinct possible outcomes of such competing effects, ranging from  superconductivity with significantly enhanced transition temperatures, to   ``naked" NFLs down to the lowest temperatures \cite{raghu2015metallic,wang2016superconductivity,damia2021thermal}.

 Real materials always host structural imperfections.  Whether such randomness can be neglected or whether they crucially determine universal properties of the system has been actively debated for decades. The  interplay between pairing and NFL behavior can significantly be altered by quenched randomness, 
which can profoundly influence the universal behavior near QCPs as well as nearby superconducting domes.    On the experimental side, the recent discovery of superconductivity in infinite-layer nickelates \cite{li2019superconductivity}, which exhibit more structural imperfections than their cuprate cousins, and which also exhibit apparent quantum critical behavior upon doping, invites us to consider quenched randomness effects on superconductivity near QCPs.   Other experimental examples where the interplay of quenched disorder and superconductivity occur include the cuprate and iron-based superconductor families, each of which possesses members with varying impurity concentrations. On the theoretical side, recent studies of the effects of quenched disorder at QCPs have attracted considerable attention \cite{MaslovQCP,nosov2020interaction,halbinger2021quenched,wu2022quantum,patel2022universal}.  Nevertheless, the degree to which superconductivity near QCPs is affected by disorder remains a largely unexplored and fundamental theoretical challenge.  One might naively expect, for instance, that disorder at QCPs might not change the pairing scale of an $s$-wave superconductor, in accordance with so-called ``Anderson's theorem" \cite{abrikosov1959theory,abrikosov1959superconducting,anderson1959theory}.




Even in conventional dirty superconducting thin films, Anderson localization \cite{anderson1958absence} significantly modifies the effective pairing vertex due to strong mesoscopic correlations of single-particle wave functions in energy and real space (so-called ``multifractality") \cite{chalker1990scaling,PhysRevB.76.235119,feigel2007eigenfunction,feigel2010fractal}. Depending on the nature of the electron-electron interaction, this effect could either substantially enhance \cite{feigel2007eigenfunction,feigel2010fractal,burmistrov2012enhancement,burmistrov2015superconductor,Mayoh2015,gastiasoro2018enhancing,Matveenko2020,fan2020superconductivity,Fan2020,Stosiek2020,burmistrov2021multifractally,andriyakhina2022multifractally,zhang2022enhanced} or suppress \cite{maekawa1982localization,takagi1982anderson,maekawa1984theory,anderson1983theory,finkel1987superconductivity,finkel1994suppression,Smith1995,PhysRevLett.83.191,kravtsov2012wonderful,antonenko2020suppression} superconductivity even at relatively weak disorder, long before a putative superconductor-insulator transition (SIT)\cite{bulaevskii1984localization,ma1985localized,kapitulnik1985anderson}.

\begin{figure}[h!]
 \center{\includegraphics[width=0.9\linewidth]{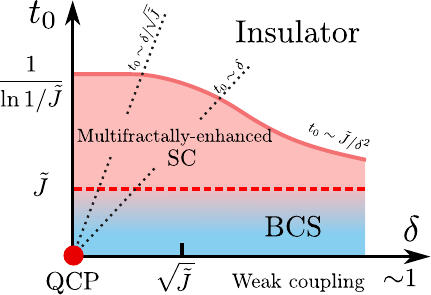}}
\caption{Schematic $T=0$ phase diagram of the model \eqref{eq:model} in terms of the detuning
from the QCP, $\delta$, and the Drude resistance, $t_0$, for $\tilde{J}\ll 1$. The solid light red line corresponds to a transition from the s-wave, spin-triplet, orbital-singlet SC state to an Anderson insulator. The dashed red line corresponds to a crossover from a ballistic regime (at weak disorder), to a dirty limit where multifractality enhances SC. The black doted lines separate Fermi-liquid and critical regimes, with a crossover region in between. Above the QCP, ferromagnetic fluctuations give rise to NFL behavior with dynamical critical exponent $z=4$ ($z=3$) in the dirty (ballistic) regime.
 }
\label{fig:main_phase_diag}
\end{figure}

The goal of the present paper
is to bridge the gap between the existing theories of NFL superconductivity in `clean' quantum critical systems, on the one side, and conventional dirty superconducting thin films, on the other.  We study the concrete problem of disordered electrons near an Ising ferromagnetic QCP. We also allow for orbital and spin degeneracy, which enables us to study the onset of $s$-wave (with respect to the Cooper pair angular momentum), spin triplet, orbital singlet  pairing near the QCP.  

More specifically, we develop a unified analytic approach to the superconducting transition in 2D weakly disordered fermionic systems coupled to a QCP, accounting for multifractality, localization, and non-Fermi liquid effects. Assuming that the mean free path is shorter than the superconducting coherence length (so-called ``dirty" limit), we derive the equation for the spectral gap function and solve it in various limits, interpolating between the Fermi-liquid and quantum critical regimes. In all cases we find that multifractal correlations significantly enhance $T_c$ in the range of intermediate disorder strength, before the SIT is eventually reached. This enhancement grows rapidly upon approach to criticality (see Fig.~\ref{fig:main_phase_diag}).

The outline of the paper is as follows. In Sec.~\ref{Sec:Model} we specify the model of a disordered Fermi liquid coupled to a quantum critical point. Next, in Sec.~\ref{Sec:Results} we present the main results of the paper. The details of our approach based on the modified Usadel equation are provided in Sec.~\ref{Sec:Usadel}.
The analysis of the Usadel equation is presented in Sec.~\ref{Sec:FL} and Sec.~\ref{Sec:NFL} in the Fermi-liquid and quantum critical regimes, respectively. Our conclusions are summarized in Sec.~\ref{Sec:Conclusions}. Some details of calculations are presented in the Appendices.

\section{Model}\label{Sec:Model}
 In the vicinity of a continuous phase transition, the system exhibits a diverging correlation length, and exhibits fluctuations on all length scales.  Among such fluctuations, the slowest modes dominate the universal properties of the system.  Indeed such reasoning underlies the Landau-Ginzburg-Wilson paradigm of critical phenomena and applies equally to metallic critical behavior at zero temperature.  Due to the diverging length scales, a microscopic model is not necessary (since many microscopic underlying models lead to the same low energy behavior near the phase transition).  Instead, one keeps the most relevant low energy degrees of freedom and analyzes their fate as the transition is approached.  In the present context, we consider ferromagnetic ordering tendency in a disordered metal, and in such a system, slow magnetic fluctuations result in a diverging magnetic correlation length and enhanced magnetic susceptibility.  The Ginzburg-Landau-Wilson theory of such a system involves a Landau Fermi liquid coupled to slow order parameter fluctuations (as determined by the magnetic susceptibility).  The coupling between fermions and magnetic fluctuations is determined entirely by symmetry\cite{landau2013statistical,chaikin_lubensky_1995, Sachdev}. Moreover, in systems with sizable spin-orbit coupling, the global spin SU(2) symmetry is explicitly broken and motivated by such systems, we consider an Ising ordering transition.  In this case, the magnetic susceptibility will be largest along an ``easy axis", which we take without loss of generality to be the $z-$direction.

More specifically, we consider a 
2D system 
of fermions at a finite density interacting via critical Ising-ferromagnetic fluctuations. The effective action $\mathcal{S}=\mathcal{S}_0+\mathcal{S}_{\rm int}$ given by
\be
\begin{aligned}\label{eq:model}
    \mathcal{S}_{0} &=  \int d\tau d\mathbf{r} \sum\limits_{\sigma b}\bar{\psi}_{\sigma b}\left[\partial_\tau +\epsilon(i\nabla) +V(\mathbf{r})\right]\psi_{\sigma b}\;, \\
   \mathcal{S}_{\rm int} &= -\frac{J}{2}\int d\tau d\mathbf{r}d\mathbf{r}' \; S^{z}(\mathbf{r},\tau) \chi^{(0)}_{zz}(\mathbf{r}-\mathbf{r}')S^{z}(\mathbf{r}',\tau)\;.
\end{aligned}
\ee
Here $\epsilon(p){=}p^2/2m -\mu$ ($m$ denotes the fermion mass and $\mu$ is the chemical potential), $S^z= \frac{1}{2}\sum_{b}\left(\bar{\psi}_{\uparrow b}\psi_{\uparrow b}-\bar{\psi}_{\downarrow b}\psi_{\downarrow b}\right)$
is the total $z$-component of the fermionic spin operator, and $J>0$ corresponds to ferromagnetic exchange. The indices $\sigma=\uparrow, \downarrow$ stand for the spin projections, and $b=1,2$ is the ``orbital" index. The bare spin susceptibility $\chi^{(0)}_{zz}(\mathbf{r})$ (or equivalently, a static paramagnon propagator) is defined through its Fourier transform $[\tilde{\chi}^{(0)}_{zz}(q)]^{-1}{=} x +c^2q^2$, and $x$ is proportional to a tuning parameter for the QCP. A common experimental tuning parameter near a ferromagnetic QCP is pressure (see for instance Ref. \cite{Brando2016}). The parameter $x$ can thus be taken to be time-reversal invariant and does not cutoff superconducting instabilities. Since the inverse susceptibility determines the quadratic coefficient of the Ginzburg-Landau expansion in powers of the order parameter, it is the most crucial parameter associated with the magnetic fluctuations. Microscopically, the interaction in Eq.~\eqref{eq:model} could emerge, for instance, after integrating out a sub-system of critical Ising spins \cite{YoniFQCP}, or as an effective contribution from high-energy degrees of freedom in a model with a short range four-fermion interaction tuned to a Stoner instability \cite{ChubukovFQCP,ChubukovMaslov}. A random potential $V(\mathbf{r})$ coupled to the fermionic density has a Gaussian distribution with the zero mean and 
a variance $\langle V(\mathbf{r})V(\mathbf{r}^\prime) \rangle = (2\pi \nu \tau)^{-1} \delta(\mathbf{r}-\mathbf{r}^\prime)$. Here $\nu$ is the density of states per spin/orbital degree of freedom, and $\tau$ is the mean free time. The spin fluctuations induce an attractive interaction in the $s$-wave, orbital-singlet, spin-triplet Cooper channel,
resulting in superconducting order 
$\psi_{\uparrow 1}\psi_{\uparrow 2}+\psi_{\downarrow 1}\psi_{\downarrow 2}$.

Our analysis of the model \eqref{eq:model} is based on several assumptions. First, we assume that the dimensionless Drude resistance per spin/orbital degree of freedom $t_0\propto 1/(\mu \tau)$ is the small expansion parameter of the model, $t_0\ll 1$. Second, we work in the `dirty' limit $\tau T_c\ll 1$, i.e. the mean free path $l=v_F\tau$ is small compared to the ballistic coherence length $v_F/T_c$, and the motion of fermions is diffusive (in the opposite limit, disorder can be ignored, and the existing results apply \cite{ChubukovFQCP, YoniFQCP}). The inverse ferromagnetic correlation length $\xi_{\rm QCP}^{-1}\equiv \sqrt{x-2\nu J}/c\ll k_F$ is used to define a dimensionless tuning parameter $\delta = ( k_F \xi_{\rm QCP})^{-1}$ allowing us to interpolate between two limits: the Fermi liquid regime, $t_0\ll \delta \ll 1$, and the quantum critical regime, 
$\delta\ll t_0\tilde{J}^{1/2}$. In the former case, the interaction is effectively short-ranged on the diffusive scales, while in the latter, it behaves as $\sim 1/q^2$ for $\sqrt{ T_c/v_F l}\ll q\ll 1/l$, inducing significant NFL effects. 

\section{Results}\label{Sec:Results}
The low-temperature behavior of the model \eqref{eq:model} is governed by the modified Usadel equation derived at the leading order in $t_0$. This equation exhibits a set of solutions, continuously varying with $\delta$, with the following features:

(i) In the Fermi-liquid regime, $t_0\ll \delta \ll 1$, the superconducting transition temperature is enhanced by multifractality in the range of parameters $\tilde{J}\ll t_0\ll \tilde{J}/ \delta^2\ll 1$, and scales as
\be \label{eq:T_c_short}
T_c \sim \tau^{-1} \exp \left\{-\delta/(t_0\tilde{J})^{1/2} \right\}\;,
\ee
where $\tilde{J}=J/(4\pi^3c^2\mu)$ is the dimensionless coupling strength. The standard BCS mechanism becomes effective only at very weak disorder $t_0\ll \tilde{J}$ (provided that $\tilde{J}\ln\omega_D\tau \ll \delta$, with $\omega_D$ serving as a UV frequency cut-off for BCS interactions), when $T_c$ crosses over to the mean-field result $T_{BCS} \sim \omega_D \exp \{-\delta/\tilde{J} \}$\footnote{The effective coupling in the Cooper channel $\lambda_{BCS}$ can be estimated from the bare spin susceptibility $\tilde{\chi}^{(0)}_{zz}(q)$ as $\lambda_{BCS}\sim \tilde{J}/\delta$, which is much smaller than the coupling in the particle-hole channel $\sim \tilde{J}/\delta^2$ due to suppression of large momentum scattering in $\tilde{\chi}^{(0)}_{zz}(q)$.}. At stronger disorder $(2\pi)^2\tilde{J}/\delta^2\lesssim t_0 \ll 1$, the superconductor-insulator transition occurs (see Fig.~\ref{fig:main_phase_diag}).

In the quantum critical regime, 
$\delta{\ll}t_0\tilde{J}^{1/2}$, we find that:

(ii) there are severe NFL self-energy effects, rendering diffusive particle-particle modes (so-called `Cooperons') strongly incoherent at scales below $\omega_{4}= t_0 J/(8c)^2{\ll}\tau^{-1}$, with anomalous dynamical scaling $z=4$.

(iii) At the same time, the pairing vertex is also enhanced by multifractality, tipping the balance in favour of superconductivity, with a power-law scaling of $T_c$:
\be \label{eq:T_c_qcp}
T_c\sim \frac{t_0J}{c^2}\left(1+\frac{t_0}{2}\ln \frac{1}{\tilde{J}}\right)\;,
\ee
where we also included the first sub-leading correction in powers of $t_0$. Remarkably, $T_c$ in Eq.~\eqref{eq:T_c_qcp} is enhanced compared to the transition temperature in the absence of disorder (which is given by a different power-law $T^{(\rm clean)}_c \sim J^2/(c^4\mu)$, see for instance \cite{abanov2020interplay})
for intermediate values of Drude resistance $\tilde{J}\ll t_0$. This behavior continues until the sub-leading correction in \eqref{eq:T_c_qcp} becomes of the order of $\mathcal{O}(1)$, i.e. for $t_0\ll 1/\ln(1/\tilde{J})$. For  stronger disorder (or equivalently, exponentially weaker coupling $\tilde{J}\ll \exp\{-2/t_0\}$) the system undergoes the localization transition (see Fig.~\ref{fig:main_phase_diag}). We also emphasize that the regime of multifractally-enhanced superconductivity broadens rapidly upon approach to the QCP.

\section{Modified Usadel equation}\label{Sec:Usadel}
\begin{figure}[t!]
 \center{\includegraphics[width=0.9\linewidth]{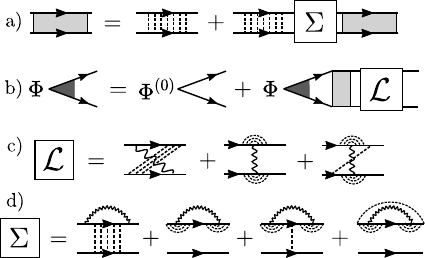}}
\caption{Diagrammatic representation of the one-loop linearized equation for the pairing vertex $\Phi_\ep$ (depicted in (b)), involving an effective pairing amplitude (shown in (c)), and the Cooperon self-energy $\Sigma_{\ep}$ (which is related to the $Z_\ep$ factor as $|\ep|Z_{\ep}\equiv |\ep|+\Sigma_\ep$). The impurity line is denoted by dashed line, and the single-particle fermionic propagator is denoted by solid line. Wavy solid line represents the dynamically screened ferromagnetic fluctuation propagator. Grey rectangular area (defined in (a)) represents the Cooperon dressed by self-energy shown in (d), as well as by weak-localization corrections to the diffusion coefficient (not shown explicitly). Some diagrams have their symmetric counterparts.}
\label{fig:main_diag}
\end{figure}

At the semiclassical (mean-field) level, properties of disordered superconductors are usually described by the Usadel equation governing the quasiclassical Green's function parameterized by the spectral angle $\theta_{\ep}$ \cite{Usadel}. In order to account for quantum corrections, we incorporate interactions of diffusive modes in the parametrically broad energy interval $T_c\ll \ep \ll 1/\tau$. This is accomplished by means of the standard fermionic perturbation theory, diagrammatically summarized in Fig.~\ref{fig:main_diag} (this approach is fairly standard in the theory of conventional disordered superconductors, see Refs.~\cite{finkel1994suppression,PhysRevLett.83.191,Skvortsov2005} for details). In addition, we also obtained the same results within the nonlinear $\sigma$-model framework, extending the approach of \cite{burmistrov2021multifractally} to the case of critical interactions (see details in Appendices \ref{Appendix1} and \ref{Appendix2}). As a result, all physical parameters of the system become scale-dependent, and the Usadel equation acquires the following form
\be \label{eq:Usadel_eq}
\frac{D}{2}\nabla^2\theta_{\ep_n}+\Phi_{\ep_n} \cos\theta_{\ep_n}-|\ep_n|Z_{\ep_n} \sin \theta_{\ep_n}  =0\;.
\ee
Here $\ep_n=\pi T (2n+1)$ denotes fermionic Matsubara frequency, and $D=v_F^2\tau/2$ is the diffusion coefficient. In striking contrast to the standard Usadel equation \cite{Usadel}, Eq.~\eqref{eq:Usadel_eq} has the energy dependent pairing vertex $\Phi_{\ep_n}$ and the self-energy factor $Z_{\ep_n}$, which are given by
\be \label{eq:Z_Phi_theta}
\begin{aligned}
\Phi_{\ep_n}&=\Phi^{(0)}+T\sum\limits_{m}\mathcal{L}_{\ep_n,\ep_{m}}\sin\theta_{\ep_m}\;,\\
Z_{\ep_n}&=1+\frac{T}{|\ep_n|}\sum\limits_{m}\operatorname{sgn}(\ep_{m})\mathcal{L}_{|\ep_n|,\ep_{m}}\cos\theta_{\ep_{m}}\;.
\end{aligned}
\ee
Here we introduced some infinitesimal `external' pairing field $\Phi^{(0)}$, which will be set to zero at the end. The effective pairing amplitude is given by
\be\label{eq:L_int}
\mathcal{L}_{\ep,\ep'} = \frac{J}{2} \int\frac{d^2q}{(2\pi)^2} \frac{\tilde{\chi}_{zz}(|\ep-\ep'|,q)}{D_{\ep,\ep'}q^2+E_{\ep}+E_{\ep'}}\;,
\ee
with $|q|\leq 1/l$. Here $E_{\ep}=|\ep|\cos\theta_{\ep}+\Phi^{(0)} \sin\theta_{\ep}$, and $[\tilde{\chi}_{zz}(|\omega_n|,q)]^{-1}=c^2(\xi_{\rm QCP}^{-2}+q^2)+\pi\nu^2J \Pi_{zz}(|\omega_n|,q)$ is the RPA-dressed spin susceptibility. 
We also allow for weak localization (WL) corrections to the diffusion coefficient $D_{\ep,\ep'}/D=1+\frac{t_0}{2}\ln[ \tau(E_{\ep}+E_{\ep'})]$ \footnote{Formally, this extension appears as a two-loop corrections to the Usadel equation at the lowest possible order in the coupling constant $\tilde{J}$, i.e. $\sim \mathcal{O}(t_0^2\tilde{J})$. Our preliminary analysis indicates \cite{two_loop} that the remaining two-loop corrections (in particular - Altshuler-Aronov-type corrections to the resistance) are at least of the order $\mathcal{O}(t_0^2\tilde{J}^2)$, and could be neglected assuming that interactions are sufficiently weak.}. 
 $\Pi_{zz}(|\omega_n|,q)$ is the dynamical part of the polarization operator and 
for arbitrary $\theta_\ep$ is given in Eq.~\eqref{eq:Pi_full_theta}. In the normal state, $\theta_{\ep}=0$, it reduces to the usual diffusive Landau damping form $\Pi_{zz}(|\omega_n|,q)=(2/\pi\nu)|\omega_n|/(Dq^2+|\omega_n|)$ \cite{MaslovQCP}, and the paramagnon propagator reads as
\be\label{eq:paramagnon_propagator}
[\tilde{\chi}_{zz}(|\omega_n|,q)]^{-1}_{\theta_{\ep}=0}=c^2(\xi_{\rm QCP}^{-2}+q^2)+ \frac{2\nu J|\omega_n|}{Dq^2+|\omega_n|}\;.
\ee

Assuming that the superconducting state is spatially homogeneous on the scale of the coherence length, the gradient term in \eqref{eq:Usadel_eq} can be ignored, and the formal solution reads as $\sin \theta_{\ep_n}= \Phi_{\ep_n}/\sqrt{(|\ep_n |Z_{\ep_n})^2+\Phi_{\ep_n}^2}$, such that Eqs.~\eqref{eq:Z_Phi_theta} become the self-consistency equations for $\Phi_{\ep_n}$ and $Z_{\ep_n}$. For the present study, we are only interested in the transition to the superconducting phase, and thus, we can assume that $\Phi_{\ep_n}$ is small, and approximate $Z_{\ep_n}$ and $\mathcal{L}_{\ep_n,\ep_m}$ by their normal state expressions. As a result, we arrive at
\be \label{eq:Full_system}
\begin{aligned}
\Phi_{\ep_n}&=\Phi^{(0)}+T\sum\limits_{m}\frac{\mathcal{L}_{\ep_n,\ep_m}}{|\ep_m| Z_{\ep_m}}\Phi_{\ep_m}\;,\\
Z_{\ep_n}&=1+\frac{T}{|\ep_n|}\sum\limits_{m}\operatorname{sgn}(\ep_m)\mathcal{L}_{|\ep_n|,\ep_m}\;,
 \end{aligned}
\ee
with the cutoff at $|\ep_m|\sim 1/\tau$, and $\mathcal{L}_{\ep_n,\ep_m}$ is now understood as the $\theta_\ep=0$ limit of Eq.~\eqref{eq:L_int}. The set of equations \eqref{eq:Full_system}, supplemented by the effective interaction \eqref{eq:L_int}, is in the core of our analysis. The diagrammatic representation of Eqs.~\eqref{eq:Full_system} is depicted in Fig.~\ref{fig:main_diag}. A few comments are in order. First, Eqs.~\eqref{eq:Full_system} were derived perturbatively in the disorder strength $t_0$, with no additional a priori assumptions on $\tilde{J}$ or $\delta$ other than that $T_c\ll 1/\tau$ (i.e. that we work in the ``dirty" limit). Therefore, all higher-order diagrams (including more complicated vertex corrections) are explicitly subleading, and could be safely ignored in our analysis of the pairing instability. This surprising simplicity stems from the interplay between {\it two} independent energy scales associated with disorder ($1/\tau$) and interactions ($J/c^2$): the latter sets the overall energy units, whereas the former is used to control perturbation theory. This is in striking contrast to the ``clean" case $T_c\gg 1/\tau$, where $J/c^2$ is the only relevant low-energy scale, and thus, no small dimensionless parameter is available. Remarkably, this implies that our theory remains under control even in the quantum critical regime $\delta=0$, with no need for any artificial small parameters (such as $1/N_F$, where $N_F$ is a large number of fermionic flavours, etc.). 
Second, Eqs.~\eqref{eq:Full_system} bare some resemblance with the standard Eliashberg equations \cite{combescot1995strong,abanov2001coherent}. However, we stress that $Z_{\ep_n}$ is not a characteristic of a single particle Green's function but rather encodes information about Green's function correlations. For similar reasons, $\mathcal{L}_{\ep_n,\ep_m}$ is not translation-invariant on the Matsubara axis, i.e. it is not a function of $|\ep_n-\ep_m|$ alone. Finally, we note that Eqs.~\eqref{eq:Full_system}, in principle, allow for a full finite temperature analysis in a parametrically broad range $T_c\ll T\ll 1/\tau$. It is known that in some cases (for instance, for a fully SU(2)-invariant QCP), the thermal self-energy effects could become significant due to the exchange of virtual bosons of zero Matsubara frequency \cite{damia2021thermal}. However, in case of the Ising symmetry, the contributions from such static modes cancel out in Eqs.~\eqref{eq:Full_system}, and do not affect the pairing instability in accordance with the Anderson's theorem  \cite{anderson1959theory, combescot1995strong}. Therefore, for the purposes of identifying the superconducting transition temperature it is sufficient to perform analytic continuation and replace Matsubara summations with continuous integrals over frequencies, in which finite temperature $T$ serves mostly as a regularization for low-energy divergences \cite{wang2016superconductivity,wang2018superconductivity,wu2020interplay}. This procedure is essentially equivalent to the $T=0$ analysis, and correctly determines the asymptotic scaling of $T_c$ with the relevant coupling constants, as well as dynamical scaling of the slow modes. The latter is guaranteed by the fact that the dynamical part of the fermionic self-energy usually dominates over the thermal one at sufficiently low temperatures. We now proceed with solving Eqs.~\eqref{eq:Full_system} in several limits.

\begin{figure*}[t!]
 \center{\includegraphics[width=1.0\textwidth]{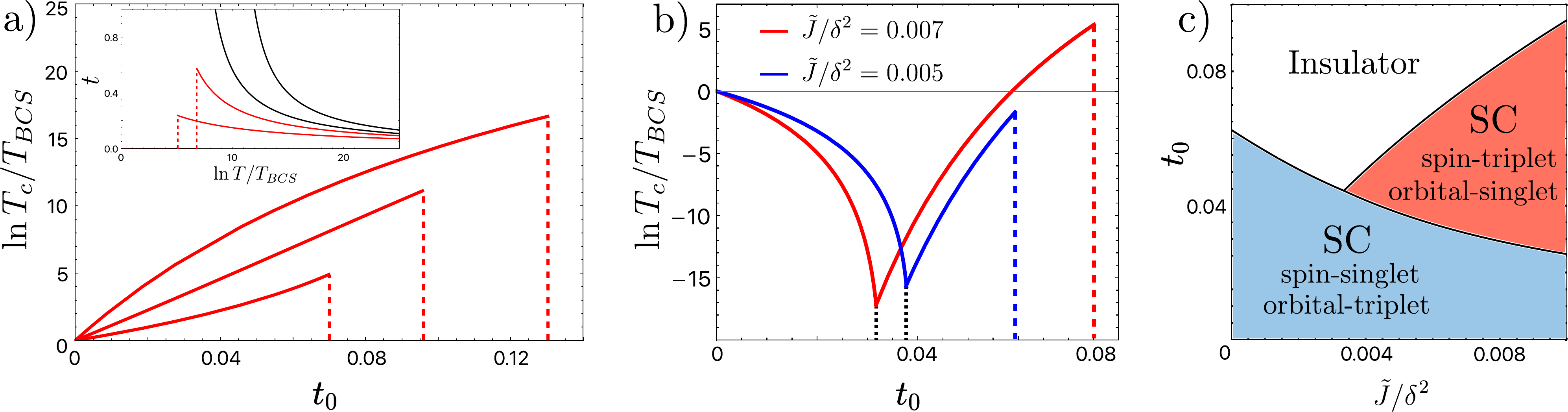}}
\caption{Superconducting transition temperature in a ``Fermi-liquid" regime ($t_0\ll\delta\ll 1$),
as a function of the bare Drude resistance $t_0$ for $\ln 1/(\tau T_{BCS})=30$. (a): the BCS coupling is in the spin-triplet channel, and $\tilde{J}/\delta^2=0.002,
\;0.005,\;0.011$ (from bottom to top). Vertical dashed lines correspond to the SIT taking place when the renormalized resistance becomes of the order of one. Inset: Temperature dependence of the renormalized resistance for $\tilde{J}/\delta^2=0.003$, $\ln 1/(\tau T_{BCS})=30$, and for $t=0.06,\;0.075,\;0.085,\;0.1$ (bottom to top). (b): The BCS coupling is in the spin-singlet channel. Black dotted lines indicate a transition between spin-singlet and spin-triplet superconducting phases, while the colored dashed lines correspond to a SIT. (c): The $T=0$ phase diagram with two distinct superconducting (SC) phases, in case of bare BCS attraction in the spin-singlet channel only, for $\ln 1/(\tau T_{BCS})=30$.}
\label{fig:main_1}
\end{figure*}

\section{Fermi-liquid regime}\label{Sec:FL}
We begin with the Fermi-liquid regime 
${t_0\ll \delta \ll 1}$, in which the interaction can be approximated as $\tilde{\chi}_{zz}(|\omega_n|,q)\approx \xi_{\rm QCP}^2/c^2$ for the entire range of momentum integration in \eqref{eq:L_int}. In a coordinate representation, this condition implies that $\chi_{zz}(r)$ decays on a scale $\xi_{\rm QCP}$ much shorter compared to the mean-free path $l$. Therefore, our model in this regime effectively describes a disordered two-orbital Fermi-liquid with strong anisotropy in the spin-exchange Landau parameters. After evaluating $\mathcal{L}_{\ep,\ep'}$, we find that $Z_\ep\equiv 1$, and the equation for the pairing vertex $\Phi_{\ep_n}$ acquires the following form
\be \label{eq:Phi_short}
\Phi_{\ep_n}=\Phi^{(0)} -2\pi \alpha T  \sum\limits_{m\geq 0} \left[\frac{\ln\tau(\ep_n{+}\ep_m)}{1{+}\frac{t_0}{2}\ln\tau(\ep_n{+}\ep_m)}\right] \frac{\Phi_{\ep_m}}{\ep_m}\;,
\ee
where $\alpha = \pi^2t_0\tilde{J}/(2\delta)^2\ll 1$ is the emergent dimensionless coupling constant. The denominator of the kernel in \eqref{eq:Phi_short} describes WL renormalization of the resistance. This equation can easily be solved with logarithmic accuracy. To this end, we approximate Matsubara sums by integrals, introduce a new variable $\zeta=\ln\frac{1}{\tau\ep}$, and replace $\ln\tau(\ep+\ep')$ by $\ln(\tau\operatorname{max}\{\ep,\ep'\})$. As a result, we find 
\be \label{eq:Gamma_1st_locality}
\Phi_\zeta =\Phi^{(0)}+\alpha \int\limits_{0}^{\zeta}d\zeta'\; f_{t_0}(\zeta')\Phi_{\zeta'}-\alpha f_{t_0}(\zeta) \int\limits_{\zeta_T}^{\zeta}d\zeta' \Phi_{\zeta'} \;,
\ee
where $f_{t_0}(\zeta) = \zeta(1-\frac{t_0\zeta}{2})^{-1}$, and $\zeta_T= \ln\frac{1}{\tau T}$. After differentiating this integral equation several times with respect to $\zeta$, it reduces to a second-order differential equation, which can be readily solved as
 \be \label{eq:Phi_solution_weak}
 \frac{\Phi_\zeta}{\Phi^{(0)}} = \sum\limits_{s=\pm} \frac{\left(1-t_0\zeta/2\right)^{-(1+s \sqrt{1{-}\eta})/2}}{1+\left(1-\frac{2}{\eta}(1+s \sqrt{1{-}\eta})\right)\left(1-\frac{t_0}{2}\zeta_T\right )}\;,
 \ee 
 where $\eta{=}16\alpha/t_0^2{\equiv} (2\pi/\delta)^2 \tilde{J}/t_0$.
 Note that $\eta$ can be arbitrary in magnitude even if both $\alpha$ and $t_0$ are small. When this ratio is smaller than $1$ (i.e. $t_0\gg \tilde{J}/\delta^2$), the superconducting susceptibility $\chi_{sc}(T)\propto\sum_{\zeta=0}^{\zeta_T}\Phi_\zeta/\Phi^{(0)}$ remains finite for all $\zeta_T<2/{t_0}$ indicating that superconductivity fails to prevent the renormalized resistance from growing uncontrollably, and thus, the ground state is insulating. In contrast, for $\eta>1$ the square root $\sqrt{1-\eta}$ becomes
 complex, meaning that superconductivity intervenes and sets in when $\chi_{sc}^{-1}(T_c){=}0$, where $T_c$ is
\be\label{eq:T_c_first}
\ln\frac{1}{\tau T_c}= \frac{2}{t_0}\left\{1-\exp\left[\frac{2(\arctan\sqrt{\eta-1}-\pi)}{\sqrt{\eta-1}}\right]\right\} .
\ee
In the limit $t_0^2\ll\alpha$, or equivalently $t_0\ll \tilde{J}/\delta^2$, Eq.~\eqref{eq:T_c_first} transforms into Eq.~\eqref{eq:T_c_short}. Crucially, the renormalized resistance at the transition remains small $t(\zeta_{T_c})=t_0/(1-t_0\zeta_{T_c}/2)\ll 1$ in this range of parameters, implying that our calculation is controllable.

We also mention in passing that, after taking the limit $t_0\rightarrow 0$ while holding $\alpha$ fixed (i.e. ignoring WL corrections to the resistivity), the superconducting susceptibilities $\chi_{sc}(T)$ acquires a particularly simple form
\be 
\lim\limits_{\substack{t_0\rightarrow 0\\
\alpha \text{ fixed}}} \chi_{sc}(T) = \frac{1}{\sqrt{\alpha}}\tan\left(\sqrt{\alpha}\zeta_T\right)\;,
\ee
which leads to the same expression for $T_c$ as in Eq.~\eqref{eq:T_c_short}. In this limit, Eq.~\eqref{eq:Phi_short} is formally analogous to the gap equation derived in \cite{Skvortsov2005,Skvortsov2012} for Coulomb repulsion. In our case, however, $\alpha$ is positive, which leads to attraction.

Another interesting limit of Eq.~\eqref{eq:Phi_solution_weak} is $\eta-1\ll 1$, which corresponds to the interactions and disorder strength being comparable in magnitude. In this case, the transition temperature takes the following form
\be \label{eq:T_c_near_critical}
T_c \sim \frac{1}{\tau} \exp \left\{-\frac{2}{t_0}\left[1-\exp\left\{-\frac{2\pi}{\sqrt{\eta-1} }+2\right\}\right]\right\}\;,
\ee
where the second term in square brackets is a small correction to the leading order estimation $T_c\sim \tau^{-1} e^{-2/t_0}$ (note that a similar scaling of $T_c$ was reported in \cite{burmistrov2012enhancement} for a different problem). In order for our analysis to be self-consistent, we also have to make sure that the resistance at this scale is still small, $t(\zeta_{T_c})\ll 1$, which leads to the following applicability condition $\ln^{-2}\frac{1}{t_0}\ll \eta-1\ll 1$.

The above expressions for $T_c$ are derived assuming that the bare large momentum BCS scattering is weak, and superconductivity emerges primarily due to attractive interaction with small momentum scattering in a particle-hole channel. In our model, this approximation is well-satisfied in a parametrically broad regime because $\tilde{\chi}^{(0)}_{zz}(q=0)\gg \tilde{\chi}^{(0)}_{zz}(q=2k_F)$ provided that $\xi_{\rm QCP} \gg k_F^{-1}$, and thus, the bare value of the corresponding dimensionless coupling constant $\lambda_{BCS}$ is suppressed by an extra small factor $\delta\ll 1$. Let us now allow for small but finite $\lambda_{BCS}$ in the spin-triplet orbital-singlet Cooper channel, which leads to a mean-field superconducting transition at some temperature $T_{BCS}\ll 1/\tau$, i.e. $1/\lambda_{BCS}=\ln \frac{1}{\tau T_{BCS}}\equiv \zeta_{T_{BCS}}\gg 1$. Then the actual transition temperature can be determined from the equation $1-\lambda_{BCS}\chi_{sc}(T_c)=0$, leading to 
\begin{equation} \label{eq:z_T_s}
    T_c\sim \tau^{-1} \exp\left\{-\frac{2}{t_0}P_{\eta}\left(\frac{t_0 \zeta_{T_{BCS}}}{2}\right)\right\}\;,
\end{equation}
where the function $P_\eta(x)$ is explicitly defined as
\be
P_\eta(x)= \begin{cases}
1-\left(\frac{x-2+x\sqrt{1-\eta }}{x-2-x\sqrt{1-\eta }}\right)^{\frac{1}{\sqrt{1-\eta }}} \;,\quad \eta<1\\ 1-\exp\left\{-\frac{2}{\sqrt{\eta-1}}\left(\pi \theta(x-2) \right.\right.\\ \quad\left. \left.-\arctan\left[\frac{x \sqrt{\eta-1}}{x-2}\right]\right)\right\}\;,\quad \eta>1
\end{cases}
\ee
The full expression in Eq.~\eqref{eq:z_T_s} reduces to either Eq.~\eqref{eq:T_c_first} or Eq.~\eqref{eq:T_c_near_critical} under the assumption $\delta^2/(\tilde{J}\ln^2\tau T_{BCS})\ll t_0 \ll \tilde{J}/\delta^2$. The resulting dependence of $T_c$ on the Drude resistance $t_0$ (as well as the temperature dependence of the renormalized resistance) is depicted in Fig.\ref{fig:main_1}(a).

It is also worth mentioning that Eq.~\eqref{eq:z_T_s}, formally continued to negative $\eta$, could be used to determine suppression of the superconducting transition temperature in the {\it spin-singlet} Cooper channel because the Ising-ferromagnetic interaction is repulsive in that channel (in this case, $T_{BCS}$ should be viewed as a mean-field transition temperature for a spin-singlet SC state). The SIT occurs when $t_0\zeta_{T_c}/2\sim 1$, which yields a critical value of the resistance $t_{0,\rm crit}\approx 4\delta^2/(\pi^2\tilde{J}\zeta_{T_{BCS}}^2)\ll 1$. Here we also assumed that $\tilde{J}\zeta_{T_{BCS}}^2\gg \delta^2 $ for the theory to remain self-consistent. We note that this critical point is analogous to the Finkelstein's result for Coulomb repulsion \cite{finkel1987superconductivity}.

In addition, one could also extend this analysis by allowing for a competing BCS attractive coupling constant in a $s$-wave {\it spin-singlet} (orbital-triplet) channel $\psi_{\uparrow 1}\psi_{\downarrow 1}+\psi_{\uparrow 2}\psi_{\downarrow 2}$, in which ferromagnetic fluctuations mediate repulsion. Under these circumstances, any increase in disorder strength will result in suppression of $T_c$ in the spin-singlet channel (see Fig.~\ref{fig:main_1}(b)), while the pairing fluctuations in the spin-triplet channel will grow. Depending on the coupling strength $\tilde{J}$, there are two possibilities (see the phase diagram in Fig.~\ref{fig:main_1}(c)): first, if the coupling is weak, the system can undergo a SIT transition directly from a spin-singlet state, and second, there could be an intermediate disorder-driven phase transition between the two distinct superconducting states, preceding a SIT (the corresponding highly non-monotonic $t_0$-dependence of $T_c$ is depicted in Fig.~\ref{fig:main_1}(b)). The latter scenario takes place provided that $T_c$ for both superconducting channels coincide, leading to the condition
\begin{align}
     \label{eq:condition_spin_signlet}
&\left(\frac{1+\sqrt{1+\eta }-4/(t_0 \zeta_{T_{BCS}})}{1-\sqrt{1+\eta }-4/(t_0 \zeta_{T_{BCS}})}\right)^{\sqrt{\frac{\eta-1}{\eta +1}}} \notag\\&=  \exp\left[-2\pi+ 2\arctan\left(\sqrt{\eta-1}\right)\right],\quad \eta>1\;.
\end{align}
This equation determined the boundary between the spin-singlet and spin-triplet SC phases in Fig.~\ref{fig:main_1}(c).
On the other hand, the onset of the SIT occurs when the renormalized resistance $t(\zeta_{T_c})$ becomes of the order of one, i.e. $t_0\simeq 1- P_{-\eta}\left(\frac{t_0\zeta_{T_{BCS}}}{2}\right)$. The latter condition sets the boundary for the insulating phase in Fig.~\ref{fig:main_1}(c). The magnitude of $T_c$ in a spin-triplet superconducting phase could become both higher or lower than the initial mean-field transition temperature $T_{BCS}$, depending on the coupling strength (both possibilities are depicted in Fig.~\ref{fig:main_1}(c)).

\section{Quantum critical regime}\label{Sec:NFL}
Next, we turn to the quantum critical regime, emerging in the limit 
$\delta\ll t_0\tilde{J}^{1/2}$ (i.e. $\xi_{\rm QCP}^{-1}$ is negligible).
In this limit, the effective pairing amplitude can be estimated as 
\be \label{eq:L_crit_2}
\mathcal{L}_{\ep,\ep'}\approx \frac{\pi}{2}  \left(\frac{\omega_{4}}{|\ep-\ep'|}\right)^{1/2}\Upsilon_{\tau(|\ep|+|\ep'|)}\;.
\ee
The square-root scaling here appears after expanding  \eqref{eq:L_int} at the lowest order in disorder strength, while preserving energy-dependence of resistance. The latter contributes a factor $\Upsilon_{x}\equiv [1+t_0 \ln(x)/2]^{-1/2}$, smoothly varying on top of the overall power-law behavior of \eqref{eq:L_crit_2}. The corresponding self-energy factor for a Cooperon reads as
\begin{align} &Z_\ep =1+\frac{\sqrt{t_0J}}{32 c } \int\limits_0^{+\infty} \frac{d\ep'}{\ep\left(1+\frac{t_0}{2}\ln\tau (\ep+\ep')\right)^{1/2}}\times\notag\\\label{eq:Z_4}&\times\left(\frac{1}{|\ep-\ep'|^{1/2}}-\frac{1}{|\ep+\ep'|^{1/2}}\right)
\approx 1+\omega_{4}^{1/2}|\ep|^{-1/2} \Upsilon_{\tau|\ep|}\;.
\end{align}
As a result, the self-consistent solution for the diffusive particle-particle propagator (`Cooperon') at the lowest order in $t_0$ takes the form 
\be \label{eq:cooperon_full}
[\mathcal{D}_{q}(\ep,\ep')]^{-1}= Dq^2+\omega_4^{1/2}(|\ep|^{1/2}+|\ep'|^{1/2})\;,
\ee
where the non-analytic frequency dependence originates from the self-energy factor $Z_{\ep_n}$ in Eq.~\eqref{eq:Z_4}, with ${\Upsilon_{\tau |\ep|}\approx 1}$ for $t_0\ll 1$.
After combining these results together, we arrive at the pairing vertex equation 
\be 
\label{eq:pair_gamma=1/2}
\Phi_{\ep_n}=\Phi^{(0)}+\frac{\pi T}{2}\sum\limits_{m\neq n} \frac{|\ep_m|^{-\frac{1}{2}}\Upsilon_{\tau (|\ep_n|+|\ep_m|)}\Phi_{\ep_m}}{|\ep_n{-}\ep_m|^{\frac{1}{2}}\left(|\frac{\ep_m}{\omega_4}|^{\frac{1}{2}}{+}\Upsilon_{\tau |\ep_m|}\right) }\;.
\ee
In solving this equation, one can make use of the fact that $\Upsilon_{\tau \ep}$ varies very smoothly compared to the power-law factor in the kernel, allowing to approximate $\Upsilon_{\tau \ep}$ by its value at $\ep=T_c$. The remaining constant factor $\Upsilon_{\tau T_c}$ can be eliminated by rescaling frequencies as $\ep \rightarrow \Upsilon_{\tau T_c}^2 \tilde{\ep}$, leaving us with a particular limit ($\gamma=1/2$) of the general gap equation extensively studied in \cite{abanov2020interplay}. Following the logic of \cite{abanov2020interplay}, we approximate $|\ep-\ep'|$ by $\operatorname{max}\{|\ep|,|\ep'|\}$ in Eq.~\eqref{eq:pair_gamma=1/2}, and reduce the resulting integral equation to a differential one, similarly to our analysis of the Fermi liquid regime \footnote{Strictly speaking, this approximation is under control only for purely logarithmic kernels (which was the case in the `Fermi liquid' regime). However, extensive studies of this approximation showed \cite{wang2018superconductivity} that it produces parametrically correct predictions for the onset of the pairing instability even in case of power-law kernels.}. It is also convenient to introduce a physical gap function $\Delta_{\ep_n} = \Phi_{\ep_n}/Z_{\ep_n}$, which obeys the following integral equation
\be \label{eq:Delta_SM_eq}
 \Delta_{\ep_n} = \Phi^{(0)} +T\sum\limits_{m}\frac{\mathcal{L}_{|\ep_n|,\ep_m}}{|\ep_m|}\left(\Delta_{\ep_m} - \frac{\ep_m}{|\ep_n|}\Delta_{\ep_n}\right)\;.
 \ee
From this representation, it is clear that the term with $\ep_m=|\ep_n|$ in the sum is canceled out. After replacing Matsubara sums with integrals, we obtain
\begin{equation}
(1+2v g_{t_0}(\xi))\Delta_\zeta =v \int\limits_{0}^{\zeta}d\zeta' \;g_{t_0}(\zeta') \Delta_{\zeta'}+v g_{t_0}(\zeta) \int\limits_{\zeta}^{\zeta_T}d\zeta' \Delta_{\zeta'}
\end{equation}
where $v=\pi \tilde{J}^{1/2}/8$,  and $g_{t_0}(\zeta)=e^{\zeta/2}(1-t_0\zeta/2)^{-1/2}$. It is instructive to compare this equation with its analog Eq.\eqref{eq:Gamma_1st_locality}, analyzed in the Fermi-liquid regime. The main conceptual distinction between these two equations, apart from having different kernels $f_{t_0}(\zeta)\neq g_{t_0}(\zeta)$, is that here we have a self-energy term, suppressing superconductivity due to NFL effects.  After differentiating this equation twice with respect to $\zeta$, and taking the limit $t_0\rightarrow 0$ (which corresponds to the absence of weak localization corrections to the resistance), we find
\be 
\left(1+2v e^{\zeta/2}\right)\ddot{\Delta}_{\zeta}+\left(v e^{\zeta/2}-\frac{1}{2}\right)\dot{\Delta}_{\zeta}+\frac{v}{2}e^{\zeta/2}\Delta_{\zeta}=0\;.
\ee
The general solution of this equation can be easily found in terms of certain hypergeometric functions. Crucially, this solution changes sign at $v \exp{\zeta/2}\sim 1$, and then starts oscillating. Indeed, for $v \exp{\zeta/2}\gg 1$ we have
\be \label{eq:Delta_c_1}
 \Delta_{\zeta}= C e^{-\zeta/4} \cos\left(\sqrt{3}\zeta/4+\phi\right)\;,
\ee
where $C$ is the overall normalization constant. The phase $\phi$ could be determined by matching Eq.~\eqref{eq:Delta_c_1} with the solution in the opposite limit, $v \exp{\zeta/2}\ll 1$, where we obtain 
\be 
 \Delta_{\zeta}= C_1 e^{\zeta/4}J_{1}\left(2\sqrt{2v}e^{\zeta/4}\right)+C_2 e^{\zeta/4}Y_{1}\left(2\sqrt{2v}e^{\zeta/4}\right)\;.
\ee
Here $C_{1,2}$ are some constants, and $J_1(x)$ and $Y_1(x)$ are the Bessel functions of the first and second kind, respectively. Therefore, the gap function indicates a pairing instability at the scale $ T_c\sim \tau^{-1}v^2\sim \tau^{-1}\tilde{J}$, which is exactly the leading term in Eq.~\eqref{eq:T_c_qcp}. In order to retain the first correction to this result, one has to expand $g_{t_0}(\zeta)$ to the lowest order in $t_0$, leading to the following equation
\begin{gather} 
\left(1+2v e^{\frac{1}{2}(1+\frac{t_0}{2})\zeta}\right)\ddot{\Delta}_{\zeta}+\left[1+\frac{ t_0}{2}\right]\left(v e^{\frac{1}{2}(1+\frac{t_0}{2})\zeta}-\frac{1}{2}\right)\dot{\Delta}_{\zeta}\notag\\\label{eq:mod_eq_SM_2}+\frac{v}{2}\left[1+\frac{ t_0}{2}\right]e^{(1+t_0/2)\zeta/2}\Delta_{\zeta}=0\;.
\end{gather}
In turn, Eq.~\eqref{eq:mod_eq_SM_2} can also be solved with hypergeometric functions. Thus, the oscillations develop when $e^{(1+t_0/2)\zeta/2}\sim v^{-1}$, leading to $T_c\sim \tau^{-1}\tilde{J}^{\frac{1}{1+t_0/2}}\approx \tau^{-1}\tilde{J}(1+\frac{t_0}{2}\ln 1/\tilde{J})$, in agreement with Eq.~\eqref{eq:T_c_qcp}.

Therefore, we find that the pairing instability in a critical regime sets in at the scale determined by Eq.~\eqref{eq:T_c_qcp}. At the same time, it is known that in the absence of disorder $T^{(\rm clean)}_c$ scales as $J^2/(c^4\mu)$ \cite{abanov2020interplay}.
The ratio of these scales is proportional to the ratio of two small dimensionless parameters $t$ and $\tilde{J}$, controlling disorder and interaction strength, respectively
\be 
T^{(\rm dirty)}_c/T^{(\rm clean)}_c \sim t_0/\tilde{J}\;.
\ee
Therefore, the conclusion is quite remarkable: even in the presence of strong NFL effects, the transition temperature is enhanced at intermediate (but still weak) disorder  $\tilde{J}\ll t_0\ll 1$. We emphasize that even though $\tilde{J}\ll 1$ in the regime of interest, this result has nothing to do with perturbation theory in $\tilde{J}$. The actual expansion is performed in powers of $t_0$ only.

The physical reason of this enhancement is twofold. At the semiclassical level, the ferromagnetic order parameter mixes with the continuum of diffusive particle-hole excitations of the Fermi surface. As a result, the effective electron-electron interaction gets Landau-overdamped, but with a dynamical scaling $z=4$ (i.e. $\omega \sim q^4$ at low energies), instead of $z=3$ as in case of ballistic dynamics. This effect by itself is already enough to produce non-analytic corrections in the Cooper channel. But most importantly, strong mesoscopic (multifractal) correlations of single-particle wave-functions (represented by the diagrams in Fig.\ref{fig:main_diag}(c)) manifest themselves in the appearance of a Cooperon in the effective pairing amplitude \eqref{eq:L_int}, further enhancing the degree of non-analyticity. In combination, multifractality and diffusive Landau damping lead to Eq.~\eqref{eq:L_crit_2}, and eventually, to the power-law scaling of $T_c$ in Eq.~\eqref{eq:T_c_qcp}.

We also note that the critical interaction remains dynamically screened at the relevant momentum scales $q\gg \sqrt{\omega/D}$ contributing to the scattering processes in Fig.\ref{fig:main_diag}(c). This precludes local mesoscopic correlations from being effectively `averaged out' at large distances, as it happens for unscreened Coulomb repulsion \cite{finkel1987superconductivity}. Moreover, despite the Cooperon propagator Eq.~\eqref{eq:cooperon_full} exhibiting anomalous dynamical scaling, the superconducting coherence length $\xi$ still obeys a standard relation to $T_c$, i.e. $\xi=\sqrt{D/T_c}$, characteristic of conventional disordered superconductors. However, the magnetic correlation length $\sim\sqrt{t_0}\xi$, as inferred from the critical paramagnon propagator Eq.~\eqref{eq:paramagnon_propagator}, appears to be parametrically shorter than $\xi$.

 At even weaker coupling (or stronger disorder), localization corrections to resistance become more noticeable, and the full frequency-dependence of $\Upsilon_{\tau\ep}$ starts to play a role, giving way to more complicated behavior of the pairing vertex $\Phi_\ep$. In particular,  the renormalized resistance evaluated at the superconducting transition temperature $t(T_c)\equiv t_0 \Upsilon_{\tau T_c}^2$ becomes of the order of one at $t_0\sim 1/\ln (1/\tilde{J})$, indicating that the system undergoes a localization transition.

\section{Conclusions}\label{Sec:Conclusions}
We have developed the theory of a pairing instability in a disordered 2D fermionic system coupled to a ferromagnetic quantum critical point. Our approach, based on the modified Usadel equation, allows to treat weak localization and non-Fermi liquid effects on equal footing, and predicts a strong enhancement of superconductivity at intermediate disorder strength, both away and near the critical point, caused by mesoscopic (``multifractal") correlations of single-particle wave-functions. In its present form, our approach does not account for phase fluctuations of the superconducting order parameter, which drive the true transition to be of Berezinskii-Kosterliz-Thouless (BKT) type. However, it is known \cite{BKT2015} that the actual transition temperature $T_{BKT}$ differs only slightly from the mean-field transition temperature $T_c$ in the limit of small resistance, and thus, our predictions are expected to remain qualitatively correct even for $T_{BKT}$.

 Our results constitute a first step towards our understanding of the fundamental interplay between disorder and superconductivity in 2D quantum critical itinerant electron systems.  Our theory and the predicted enhancement of superconductivity could in principle be tested using sign problem free Monte Carlo simulations of metallic criticality along the lines of Ref. \cite{YoniFQCP}. In the future, we wish to investigate properties of the emerging superconducting phase, where the full non-linear form of the Usadel equation \eqref{eq:Usadel_eq} will be required. Particularly intriguing observables include mesoscopic fluctuations of the local density of states \cite{burmistrov2021multifractally}, and the superfluid stiffness \cite{BKT2015}. In addition, it would be interesting to explore other types of QCPs, including cases where the critical order parameter is not conserved \cite{Oganesyan2001,lederer2015enhancement,lederer2017superconductivity}, and thus, couples differently to diffusive modes compared to the present case.

\begin{acknowledgements}
	We thank A.~Chubukov and V.~Kravtsov for fruitful discussions. The work
of PAN and SR was supported in part by the US Department of Energy, Office of Basic Energy Sciences, Division of Materials Sciences and Engineering, under contract number DE-AC02-76SF00515. ISB was partially supported by the Russian Ministry of Science and 
Higher Education and by  the Basic Research Program of HSE.
\end{acknowledgements}

 \appendix

 \section{$\sigma$-model}\label{Appendix1}
In this section we provide details of the $\sigma$-model approach to the problem of dirty superconductivity in quantum critical systems. On the technical side, our analysis extends the methodology suggested for a different problem in \cite{burmistrov2021multifractally} by allowing for arbitrary coupling strength, as well as frequency and momentum dependence of the interaction in the one-loop derivation of the effective action. As a first step, we follow the standard procedure: we average over disorder using the replica trick, decouple the resulting interaction via a matrix-valued Hubbard–Stratonovich field $Q$, and integrate out the fermions. The resulting integral over $Q$ is computed via the saddle-point method justified in the limit of weak disorder. Finally, we obtain the action for the low-energy modes (so-called `diffusons' and `cooperons') $S_\sigma+S_{\rm int}$, where
\be\label{eq:NLSM}
\begin{aligned}
S_\sigma=&\frac{\pi\nu }{8}\int d\mathbf{r} \operatorname{Tr}\left[ D(\nabla Q)^2-4 \hat{\varepsilon} Q - 4 \hat{\Phi}^{(0)} Q \right]\;,\\
S_{{\rm int}}=& \left(\frac{\pi\nu}{4}\right)^2 J T\sum\limits_{n\gamma }\sum\limits_{r=0,3}\int d\mathbf{r}d\mathbf{r}'\; \chi_{zz}^{(0)}(|\omega_n|,\mathbf{r}-\mathbf{r}')\\
&\times\operatorname{Tr}\left[ I^\gamma_n t_{r30}Q(\mathbf{r})\right] \operatorname{Tr}\left[ I^\gamma_{-n} t_{r30}Q(\mathbf{r}')\right]\;,
\end{aligned}
\ee
where $Q$ are matrices operating in the replica, Matsubara, particle-hole, spin, and orbital spaces. The interaction amplitude is given by
\be
[\tilde{\chi}_{zz}^{(0)}(|\omega_n|,q)]^{-1}=c^2(\xi_{\rm QCP}^{-2} +q^2)\;.
\ee
 The inverse screening length $\xi_{\rm QCP}^{-1}=\sqrt{x-2\nu J}/c$ describes the deviation from the quantum critical point (QCP). Here $x$ is the tuning parameter, and $2\nu J$ comes from the static polarization operator. In principle, the initial electron-electron interaction of our model (Eq.\eqref{eq:model} of the main text) also leads to a number of additional operators in $S_{{\rm int}}$ representing inelastic scattering with large momentum transfer $\sim 2k_F$. However, in a vicinity of a QCP, the bare interaction is sharply picked at $q=0$, and all other bare scattering amplitudes are suppressed by extra small factors $\sim (k_F\xi_{\rm QCP})^{-1}\ll 1$ compared to the term that we retain in $S_{{\rm int}}$.

The definitions of all the matrices in \eqref{eq:NLSM} are standard:
\be 
\begin{aligned}
\left(\hat{\varepsilon}\right)_{n n^{\prime}}^{\alpha \beta} &= \ep_{n} \delta_{n n^{\prime}} \delta^{\alpha \beta}t_{000},\;\; (\hat{\Phi}^{(0)})_{n n^{\prime}}^{\alpha \beta}\equiv \Phi^{(0)} \delta_{n, -n^{\prime}} \delta^{\alpha \beta} t_{122},\\
\left(I_{k}^{\gamma}\right)_{n n^{\prime}}^{\alpha \beta}&=\delta_{n-n^{\prime}, k} \delta^{\alpha \beta} \delta^{\alpha \gamma} t_{000}\;,
\end{aligned}
\ee
where $\Phi^{(0)}$ represents an external infinitesimal pairing potential in the $s$-wave, orbital-singlet, spin-triplet Cooper channel, which we introduce for future convenience. The Greek indices $\alpha,\beta=1,...,N_r$ stand for different replicas.  The generators $t_{rsB}$ span the particle-hole/spin/orbital matrix space, and are defined as
\be 
t_{rsB}= \tau_{r}\otimes\sigma_{s}\otimes T_B, \quad r,s,B=0,1,2,3\;.
\ee
Here $\tau_{i}/\sigma_i/T_i$, $i=0,1,2,3$, are the standard Pauli matrices. The $Q$ matrix field obeys the following constraints and charge-conjugation symmetry
\begin{gather}
Q^{2}=1, \quad \operatorname{Tr} Q=0, \quad Q^{\dagger}=C^{T} Q^{T} C\;,\notag \\ \label{eq:Q_sym}
C=i\tau_1\otimes \sigma_2\otimes T_0, ,\quad C^T = - C\;.
\end{gather}

\subsection{Superconducting saddle point}
 In order to investigate properties of the dirty superconducting state, one has to look for the following saddle point structure
 \be 
\begin{aligned}\label{eq:saddle-point}
 \underbar{Q}_{nm}^{\alpha \beta}&=\left(t_{000}\cos \theta_{\varepsilon_n}\operatorname{sgn}\varepsilon_n \delta_{\varepsilon_n \varepsilon_m}\right. \\&+\left. t_{122} \sin \theta_{\varepsilon_{n}}\delta_{\varepsilon_n,-\varepsilon_m}\right)\delta^{\alpha \beta}\;,
 \end{aligned}
 \ee
which is parameterised by a yet unknown function $\theta_{\ep}$ called the `spectral angle'. The dependence of $\theta_{\ep}$ on $\ep$ and other parameters will be determined from the minimization of the resulting effective action. The off-diagonal matrix elements in \eqref{eq:saddle-point} are encoded into the generator $t_{122}$ corresponding to our particular choice of the superconducting order parameter. One can easily check that $\frac{1}{8}\operatorname{tr}t_{122}t^*_{122}=1$, and $-\frac{1}{8}\operatorname{tr}t_{122}Ct_{122}^T C=1$. It is also useful to rewrite $\underbar{Q}_{nm}^{\alpha \beta}$ as $R^{-1}\Lambda R$, where 
\be 
R_{mk}^{\alpha \beta}= \left(t_{000}\cos\theta_{\varepsilon_k}\delta_{\ep_k\ep_m}+t_{122}\sin\theta_{\ep_k}\delta_{\ep_k,-\ep_m}\right)\delta^{\alpha \beta}\;,
\ee
and $\left(\Lambda\right)_{n n^{\prime}}^{\alpha \beta}= \operatorname{sgn}\ep_{n} \delta_{n n^{\prime}} \delta_{\alpha \beta}t_{000}$ is the usual metallic saddle point used in most previous studies. Clearly, $\underbar{Q}|_{\theta_\ep\equiv 0}\equiv \Lambda$. Note also that $R^{-1}=R^\dagger$, and $CT^T=R^{-1}C$.
The $\sigma$-model action $S_\sigma$ evaluated at the saddle point $\underbar{Q}$ reads as
\be \label{eq:S_cl}
S_{\rm cl}[\theta]=-8\pi\nu N_r V \sum\limits_{n>0}\left[\Phi^{(0)} \sin\theta_{\ep_n}+ \ep_n\cos\theta_{\varepsilon_n}  \right]\;,
\ee
where $V=\int d\mathbf{r}$ is the total volume.  After differentiating the action with respect to $\theta_\ep$, we obtain the classical Usadel equation
\be \label{eq:cl_Usadel}
-|\ep_n| \sin \theta_{\ep_n} +\Phi^{(0)} \cos\theta_{\ep_n} =0\;.
\ee
 In principle, the spectral angle could slowly vary in the coordinate space as well, which would lead to a gradient term, see Eq.\eqref{eq:Usadel_eq} of the main text. However, we are interested in the superconducting state which is spatially homogeneous on the scale of the coherence length, so the gradients term in \eqref{eq:Usadel_eq} can be ignored.

The goal of the next sections is to compute interaction-induced loop corrections to the classical action \eqref{eq:S_cl}.
\begin{widetext}
\subsection{Fluctuations around the saddle point}
In order to resolve the non-linear constraints \eqref{eq:Q_sym} in combination with the non-trivial structure of the saddle point \eqref{eq:saddle-point}, the matrix field $Q$ could parameterized as
\be \label{eq:W_parametrization}
Q=R^{-1}(W+\Lambda \sqrt{1-W^{2}})R, \quad W_{\varepsilon\varepsilon'}=w_{\varepsilon\varepsilon'} \theta(\varepsilon)\theta(-\varepsilon')+\bar{w}_{\varepsilon\varepsilon'} \theta(-\varepsilon)\theta(\varepsilon')\;,
\ee
where we explicitly emphasized the structure of $W$ in the Matsubara space. Also, in everything that follows, we implement a short-hand notation for Matsubara frequencies: instead of writing the full form $\ep_{n_1}$, $\ep_{n'}$, etc., we will simply use $\ep_{1}$, $\ep'$ and so on. The blocks $\bar{w}$ and $w$ are matrices in both the replica and $t_{rsB}$-spanned spaces. They obey certain symmetry constraints
\be 
\bar{w}=-C w^{T} C, \quad w=-C w^{*} C\;.
\ee
 We decompose all fields in terms of generators $t_{rsB}$ as
\be\label{eq:w_as_generators}
 \left[w(x)\right]_{\varepsilon \varepsilon'}^{\alpha \beta} = \sum_{rsB} \left[w_{rsB}(x)\right]_{\varepsilon \varepsilon'}^{\alpha \beta} t_{rsB}\;,\quad 
 \left[\bar{w}(x)\right]_{\varepsilon \varepsilon'}^{\alpha \beta} = \sum_{rsB} \left[\bar{w}_{rsB}(x)\right]_{\varepsilon \varepsilon'}^{\alpha \beta} t_{rsB}=-\sum_{rsB} \left[w_{rsB}(x)\right]_{\varepsilon' \varepsilon}^{\beta\alpha } Ct^T_{rsB}C\;.
\ee
We also note that $
-Ct^T_{rsB}C = m_{rsB}t_{rsB}$,
where
\be \label{eq:mrsB}
m_{rsB}=(\delta_{r\neq 3}-\delta_{r3})(\delta_{s0}-\delta_{s\neq 0})(\delta_{B\neq 2}-\delta_{B2})\;.
\ee
 Therefore, the the fields $\bar{w}$ and $w$ are not independent of each other: $\left[\bar{w}_{rsB}(x)\right]_{\varepsilon \varepsilon'}^{\alpha \beta} = m_{rsB}\left[w_{rsB}(x)\right]_{\varepsilon' \varepsilon}^{\beta\alpha}$.

Next, we substitute \eqref{eq:W_parametrization} into the action \eqref{eq:NLSM} and expand up to a quadratic order in fluctuations $W$. We find 
\be
\begin{aligned}
    S^{(2)}_{\sigma }[\theta,W]&=\frac{\pi\nu}{8}D \int d \mathbf{r} \operatorname{Tr}\left(\nabla W\right)^2+\frac{\pi\nu }{4}\int d \mathbf{r} \operatorname{Tr}\left[ \hat{E} \Lambda W^2\right]\;,\\
    S^{(2)}_{\rm int}[\theta,W] &= \frac{1 }{16}(\pi\nu)^2J T\sum\limits_{n\gamma }\sum\limits_{r=0,3}\int d \mathbf{r}d \mathbf{r}'\; \chi_{zz}^{(0)}(|\omega_n|, \mathbf{r}- \mathbf{r}')\operatorname{Tr}\left[R I^\gamma_n t_{r30}R^{-1} W( \mathbf{r})\right]\operatorname{Tr}\left[R I^\gamma_n t_{r30}R^{-1}W( \mathbf{r}')\right]\;,
\end{aligned}
\ee
where $\hat{E}=R(\hat{\ep}+\hat{\Phi}^{(0)} )R^{-1}$.
Note that the term linear in $W$ vanishes 
due to the classical Usadel equation \eqref{eq:cl_Usadel}. The explicit form of the higher order terms will be given later. Finally, we compute the effective action $S_{\rm eff}[\theta]$ as
\begin{gather} 
S_{\rm eff}[\theta]=  \label{eq:effective_action}
 -\ln \int DW e^{-S_{\rm \sigma}^{(2)}[\theta,W]-S_{\rm int}^{(2)}[\theta,W]}= \int_0^{J} \frac{dJ}{J} \langle S_{\rm int}^{(2)}[\theta,W]\rangle\;,
\end{gather}
where the average in the last expression is performed with respect to the quadratic action for Gaussian fluctuations. In the next section, we compute the propagator of these fluctuations. 

\subsection{Gaussian action for fluctuations}
One can easily verify that only the modes $(030),(112),(330),(212)$ are affected by interactions at the Gaussian level. The full quadratic action has the form
\be \label{eq:Gaussian_action_1}
S_{\rm \sigma}^{(2)}[\theta,W]+S_{\rm int}^{(2)}[\theta,W]=2\pi \nu \int_q \sum\limits_{\{\varepsilon_i>0\}}\sum\limits_{r=0,3}\sum\limits_{j,k=1}^2 \sum\limits_{\{\alpha_i\}}\Psi^{\alpha_1\alpha_2,(r)}_{\ep_1,-\ep_2,k}(q)[\hat{A}_r(q)]^{\alpha_1\alpha_4;\alpha_2\alpha_3}_{\ep_1\ep_4;\ep_2\ep_3;kj}(1-2\delta_{j1}\delta_{r0})\Psi^{\alpha_4\alpha_3,(r)}_{\ep_4,-\ep_3,j}(-q)+\dots\;,
\ee
where we omitted the terms describing the modes  unaffected by interactions. We also used $\bar{\Psi}^{\alpha_1\alpha_2,(r)}_{-\ep,\ep',j}(q)=(1-2\delta_{j1}\delta_{r0})\Psi^{\alpha_2\alpha_1,(r)}_{\ep',-\ep,j}(q)$. The matrix elements of $\hat{A}_r(q)$ are given by
\begin{gather}
[\hat{A}_r(q)]^{\alpha_1\alpha_4;\alpha_2\alpha_3}_{\ep_1\ep_4;\ep_2\ep_3;kj}=\left[Dq^2 +E_{\ep_1}+E_{\ep_2} \right]\delta_{\ep_2\ep_3}\delta_{\ep_1\ep_4}\delta_{kj}\delta_{\alpha_1\alpha_4}\delta_{\alpha_2\alpha_3}\notag\\ \label{eq:Aa_def}
+2\pi \nu J T\sum\limits_n \tilde{\chi}^{(0)}_{zz}(|\omega_n|,q) 
X^{(r)}_{n,k}(\ep_1,\ep_2)[X^{(r)}_{n,j}(\ep_4,\ep_3)]^*\delta_{\alpha_1\alpha_4}\delta_{\alpha_2\alpha_3}\delta_{\alpha_1\alpha_2} \;,
\end{gather}
where $E_\ep=\ep\cos\theta_{\ep}+\Phi^{(0)}\sin\theta_{\ep}$. We also defined the following vector combinations
\be 
\begin{aligned}
   \mathbf{\Psi}_{\ep_1,-\ep_2}^{\alpha\beta,(0)}&=\left([w_{030}]_{\ep_1,-\ep_2}^{\alpha\beta},\; [w_{112}]_{\ep_1,-\ep_2}^{\alpha\beta}\right)\;\quad \mathbf{\Psi}_{\ep_1,-\ep_2}^{\alpha\beta,(3)}=\left([w_{330}]_{\ep_1,-\ep_2}^{\alpha\beta},\; [w_{212}]_{\ep_1,-\ep_2}^{\alpha\beta}\right),\\
     \bar{\mathbf{\Psi}}_{-\ep_3,\ep_4}^{\alpha\beta,(0)}&=\left([\bar{w}_{030}]_{-\ep_3,\ep_4}^{\alpha\beta},\; [\bar{w}_{112}]_{-\ep_3,\ep_4}^{\alpha\beta}\right)\;\quad  \bar{\mathbf{\Psi}}_{-\ep_3,\ep_4}^{\alpha\beta,(3)}=\left([\bar{w}_{330}]_{-\ep_3,\ep_4}^{\alpha\beta},\; [\bar{w}_{212}]_{-\ep_3,\ep_4}^{\alpha\beta}\right)\;,
\end{aligned}
\ee
where all $\ep_i>0$, and auxiliary vectors $\mathbf{X}$ are defined as 
\be 
\begin{aligned}
    \mathbf{X}_n^{(0)}(\ep_1,\ep_2) &= \left(\cos\left(\frac{\theta_{\ep_1}+\theta_{\ep_2}}{2}\right)(\delta_{\ep_1+\ep_2,-\omega_n}-\delta_{\ep_1+\ep_2,\omega_n})\;,\; -i\sin\left(\frac{\theta_{\ep_1}-\theta_{\ep_2}}{2}\right)(\delta_{\ep_1-\ep_2,\omega_n}-\delta_{\ep_1-\ep_2,-\omega_n})\right)\;,\\
     \mathbf{X}_n^{(3)}(\ep_1,\ep_2) &= \left(\cos\left(\frac{\theta_{\ep_1}+\theta_{\ep_2}}{2}\right)(\delta_{\ep_1+\ep_2,-\omega_n}+\delta_{\ep_1+\ep_2,\omega_n})\;,\; \sin\left(\frac{\theta_{\ep_1}-\theta_{\ep_2}}{2}\right)(\delta_{\ep_1-\ep_2,\omega_n}+\delta_{\ep_1-\ep_2,-\omega_n})\right)\;.
\end{aligned}
\ee
We note that $\mathbf{X}_{-n}^{(r)}=-(-1)^r\mathbf{X}_{n}^{(r)}$. After inverting the matrix in \eqref{eq:Gaussian_action_1}, we obtain the following correlation function for $\Psi$ field
\begin{gather}
    \langle\Psi^{\alpha_1\alpha_2,(r)}_{\ep_1,-\ep_2,k}(q)\bar{\Psi}^{\alpha_3\alpha_4,(r)}_{-\ep_3,\ep_4,j}(-q)\rangle= \frac{1}{4\pi\nu}\mathcal{D}^{(0)}_{q}(\ep_1,\ep_2)\delta_{\alpha_1\alpha_4}\delta_{\alpha_2\alpha_3}\Bigg\{ \delta_{\ep_1\ep_4}\delta_{\ep_2\ep_3}\delta_{jk}  -2\pi\nu J  \delta_{\alpha_1\alpha_2} \mathcal{D}^{(0)}_{q}(\ep_3,\ep_4)\notag \\\label{eq:corr_function_Psi}\left. \times T\sum\limits_{m}\tilde{\chi}_{zz}(|\omega_m|,q)\left[X_{m;k}^{(r)}(\ep_1,\ep_2)\right]^*X_{m;j}^{(r)}(\ep_4,\ep_3)\right\}\;.
\end{gather}
Here the bare diffusive propagator $\mathcal{D}^{(0)}_{q}(\ep,\ep')$ and the RPA-dressed interaction amplitude $\tilde{\chi}_{zz}(|\omega_n|,q)$ are given by
\be
[\mathcal{D}^{(0)}_{q}(\ep,\ep')]^{-1}= Dq^2+E_{\ep} + E_{\ep'}\;,\quad [\tilde{\chi}_{zz}(|\omega_n|,q)]^{-1}=c^2(\xi_{\rm QCP}^{-2}+q^2)+\pi\nu^2 J\Pi_{zz}(|\omega_n|,q)\;.
\ee
Here, $\Pi_{zz}(|\omega_n|,q)$ is the dynamical part of the spin correlation function
\be \label{eq:Pi_full_theta}
\Pi_{zz}(|\omega_n|,q)=\frac{4T}{\nu} \sum\limits_{\ep,\ep'>0}\mathcal{D}^{(0)}_{q}(\ep,\ep')\Big[\cos^2 \left(\frac{\theta_{\ep}+\theta_{\ep'}}{2}\right) \delta_{|\ep+\ep'|,|\omega_n|}+
 \sin^2 \left(\frac{\theta_{\ep}-\theta_{\ep'}}{2}\right) \delta_{|\ep-\ep'|,|\omega_n|}\Big]\;.
\ee
In the normal state, $\theta_\ep\equiv0$, this expression reduces to the standard diffusive Landau damping
\be 
\left.\Pi_{zz}(|\omega_n|,q)\right|_{\theta_\ep\equiv 0} = \left(\frac{2}{\pi\nu}\right) \frac{|\omega_n|}{Dq^2+|\omega_n|}\;.
\ee
In the next section, we use the correlation function \eqref{eq:corr_function_Psi} to compute one-loop corrections to the effective action.

\section{One-loop correction to the Usadel equation}\label{Appendix2}
According to \eqref{eq:effective_action}, in order to compute the effective action, one has to first evaluate the following average
\begin{gather}
    \left\langle S_{\rm int}^{(2)}[\theta,W]\right\rangle = 4(\pi\nu)^2 J T N_r \int_q \sum\limits_{\{\varepsilon_i>0\}}\sum\limits_{r=0,3}\sum\limits_{j,k=1}^2 \langle\Psi^{\alpha\alpha,(r)}_{\ep_1,-\ep_2,k}(q)\bar{\Psi}^{\alpha\alpha,(r)}_{-\ep_3,\ep_4,j}(-q)\rangle \sum\limits_n \tilde{\chi}_{zz}(|\omega_n|,q)
X^{(r)}_{n,k}(\ep_1,\ep_2)[X^{(r)}_{n,j}(\ep_4,\ep_3)]^* \notag\\=
 \frac{N_rV}{2} \int_q \sum\limits_{n}\frac{\pi\nu^2 J\Pi_{zz}(|\omega_n|,q)}{[\tilde{\chi}^{(0)}_{zz}(|\omega_n|,q)]^{-1}+\pi\nu^2 J\Pi_{zz}(|\omega_n|,q)}\;.
\end{gather}
After dividing this expression by $J$ and integrating over $J$ from $0$ to $J$, we obtain the following result for the effective action
\be \label{eq:S_cl+fl}
S_{\rm eff}[\theta]=-8\pi\nu N_r V \left[\sum\limits_{\varepsilon>0}\left(\Phi^{(0)} \sin\theta_\ep+ \ep\cos\theta_\varepsilon\right)  - \frac{1}{16\pi \nu}\int_q \sum\limits_n\ln \Big(1+\pi\nu^2 J \tilde{\chi}^{(0)}_{zz}(|\omega_n|,q)\Pi_{zz}(|\omega_n|,q)\Big) \right]\;.
\ee
One can now vary this action with respect to $\theta_\ep$. For this, the following identity is useful
\begin{gather}
    \frac{\delta}{\delta \theta_\ep}\Pi_{zz}(|\omega_n|,q) =-\frac{4T}{\nu}\sum\limits_{\ep'>0}\mathcal{D}^{(0)}_{q}(\ep,\ep')\Big[ \sin\theta_\ep\cos\theta_{\ep'}(\delta_{\ep+\ep',|\omega_n|}-\delta_{|\ep-\ep'|,|\omega_n|}) 
    +\cos\theta_\ep\sin\theta_{\ep'}(\delta_{\ep+\ep',|\omega_n|}+\delta_{|\ep-\ep'|,|\omega_n|})\Big]\notag\\ \label{eq:derivative_pi}- \frac{8T}{\nu}\left[\Phi^{(0)} \cos\theta_\ep-|\ep|\sin\theta_{\ep}\right]\sum\limits_{\ep'>0}[\mathcal{D}_q^{(0)}(\ep,\ep')]^2 \left[\cos^2\left(\frac{\theta_\ep+\theta_{\ep'}}{2}\right)\delta_{|\omega_n|,|\ep+\ep'|}+\sin^2\left(\frac{\theta_\ep-\theta_{\ep'}}{2}\right)\delta_{|\omega_n|,|\ep-\ep'|}\right]\;.
\end{gather}

 At the one-loop level, the modified saddle point equation for the spectral angle has the following form
\be \label{eq:mod_Usadel_1st}
F_{\ep}\left(\Phi^{(0)} \cos\theta_\ep-|\ep| \sin \theta_\ep \right) +\Phi_\ep \cos\theta_\ep-|\ep|Z_{\ep} \sin \theta_\ep =0\;,
\ee
and the expressions for $Z_{\ep}$ and $\Phi_\ep$ are given as (compare with Eq.\eqref{eq:Z_Phi_theta} of the main text)
\begin{gather}\label{eq:Z_ep}
Z_{\ep}=1-\frac{JT}{2|\ep|}\sum\limits_{\ep'>0}\int\frac{d^2q}{(2\pi)^2}\mathcal{D}^{(0)}_{q}(\ep,\ep') \left[\tilde{\chi}_{zz}(|\ep+\ep'|,q)-\tilde{\chi}_{zz}(|\ep-\ep'|,q)\right]\cos\theta_{\ep'}=1+\frac{T}{|\ep|}\sum\limits_{\ep} \operatorname{sgn}(\ep')\mathcal{L}^{(0)}_{\ep,\ep'}\cos\theta_{\ep'}\;,\\\label{eq:Deltaep}
\Phi_{\ep}=\Phi^{(0)}+\frac{JT}{2}\sum\limits_{\ep'>0}\int\frac{d^2q}{(2\pi)^2}\mathcal{D}^{(0)}_{q}(\ep,\ep') \left[\tilde{\chi}_{zz}(|\ep+\ep'|,q)+\tilde{\chi}_{zz}(|\ep-\ep'|,q)\right]\sin\theta_{\ep'}=\Phi^{(0)}+T\sum\limits_{\ep} \mathcal{L}^{(0)}_{\ep,\ep'}\sin\theta_{\ep'}\;,\\
F_{\ep}\label{eq:F_ep}
 = JT\sum\limits_{\ep'>0}\int\frac{d^2q}{(2\pi)^2}[\mathcal{D}^{(0)}_{q}(\ep,\ep')]^2\left[\cos^2\left(\frac{\theta_\ep+\theta_{\ep'}}{2}\right)\tilde{\chi}_{zz}(|\ep+\ep'|,q)+\sin^2\left(\frac{\theta_\ep-\theta_{\ep'}}{2}\right)\tilde{\chi}_{zz}(|\ep-\ep'|,q)\right]\;,
\end{gather}
for $\ep>0$, $[\tilde{\chi}_{zz}(|\omega_n|,q)]^{-1}=c^2(\xi_{\rm QCP}^{-2}+q^2)+\pi\nu^2J \Pi_{zz}(|\omega_n|,q)$, and $\mathcal{L}^{(0)}_{\ep,\ep'}$ differs from $\mathcal{L}_{\ep,\ep'}$ in Eq.\eqref{eq:L_int} of the main text by the absence of WL corrections to the diffusion coefficient. Formally, in order to promote $\mathcal{L}^{(0)}_{\ep,\ep'}$ to $\mathcal{L}_{\ep,\ep'}$, one has to compute two-loop corrections to the effective action. This calculation is much more involved than the one-loop derivation presented here, and the details will be reported elsewhere \cite{two_loop}. Our preliminary analysis of the two-loop corrections indicates that the weak localization correction $D\rightarrow D_{\ep,\ep'}\equiv D\left[1+\frac{t_0}{2}\ln\tau(E_\ep+E_{\ep'})\right]$ in the diffusive propagator (i.e. $[\mathcal{D}^{(0)}_{q}(\ep,\ep')]^{-1} \rightarrow [\mathcal{D}_{q}(\ep,\ep')]^{-1}\equiv D_{\ep,\ep'}q^2+E_{\ep} + E_{\ep'}$) is the only contribution appearing at the lowest possible order in the coupling constant $\tilde{J}$, i.e. at the order $\sim \mathcal{O}(t_0^2\tilde{J})$. Therefore, other two-loop corrections could be neglected assuming that interactions are sufficiently weak (this limit is particularly relevant for multifractality-induced effects \cite{burmistrov2012enhancement}).

In addition, we emphasize that $F_\ep$ in Eq.\eqref{eq:mod_Usadel_1st} appears from the variation of the spectral angle entering the diffusion propagator (see the last line in Eq.\eqref{eq:derivative_pi}). This contribution is accompanied by an extra factors $\Phi^{(0)} \cos\theta-\ep \sin\theta$ which coincides with the classical Usadel equation \eqref{eq:cl_Usadel}. This means that the effect of $F_\ep$ is always of the higher order in $t_0$ compared to the remaining terms. In addition, $F_\ep$ always contains `weaker' non-analyticities compared to $\Phi_\ep$ and $Z_\ep$. For instance, its perturbative effect on $T_c$ in case of Coulomb repulsion is known \cite{maekawa1982localization} to be sub-leading (of the order $\mathcal{O}(t_0\log^2\tau T_c)$) compared to the correction from $\Phi_\ep$ (which is of the order $\mathcal{O}(t_0\log^3\tau T_c)$). To summarize, we can ignore $F_\ep$ in Eq.~\eqref{eq:mod_Usadel_1st}  and reduce it to the form \eqref{eq:Usadel_eq} given in the main text. It is also convenient to re-write Eq.~\eqref{eq:Usadel_eq} as a system of two coupled equations involving $\Phi_\ep$ and $Z_\ep$ as independent functions. This can be accomplished by means of the following formal solution $\sin\theta_{\ep}=\Phi_\ep/\sqrt{(|\ep|Z_\ep)^2+\Phi_\ep^2}$, which leads to Eqs.~\ref{eq:Full_system} of the main text.

\end{widetext}
		\bibliography{NFL_resub_new}

  \end{document}